\documentclass{article}

\usepackage{microtype}
\usepackage{graphicx}
\usepackage{subfigure}
\usepackage{booktabs} % for professional tables

\usepackage{hyperref}

\usepackage[accepted]{icml2020}

\icmltitlerunning{Interpretable, Multidimensional, Multimodal Anomaly Detection with Negative Sampling}

\begin{document}

\twocolumn[
\icmltitle{Interpretable, Multidimensional, Multimodal Anomaly Detection \\
    with Negative Sampling for Detection of Device Failure}



\icmlsetsymbol{equal}{*}

\begin{icmlauthorlist}
\icmlauthor{John Sipple}{goo}
\end{icmlauthorlist}

\icmlaffiliation{goo}{Google, Mountain View, California, USA}

\icmlcorrespondingauthor{John Sipple}{sipple@google.com}

\icmlkeywords{anomaly detection, fault detection, negative sampling, anomaly, concentration, integrated gradients, 
interpretation, interpretability, smart buildings, IoT}

\vskip 0.3in
]



\printAffiliationsAndNotice{}  % leave blank if no need to mention equal contribution

\begin{abstract}
In this paper we propose a scalable, unsupervised approach for detecting anomalies in the Internet of Things (IoT). Complex devices are connected daily and eagerly generate vast streams of multidimensional telemetry. These devices often operate in distinct modes based on external conditions (day/night, occupied/vacant, etc.), and to prevent complete or partial system outage, we would like to recognize as early as possible when these devices begin to operate outside the normal modes. We propose an unsupervised anomaly detection method that creates a negative sample from the positive, observed sample, and trains a classifier to distinguish between positive and negative samples. Using the Concentration Phenomenon, we explain why such a classifier ought to establish suitable decision boundaries between normal and anomalous regions, and show how Integrated Gradients can attribute the anomaly to specific dimensions within the anomalous state vector. We have demonstrated that negative sampling with random forest or neural network classifiers yield significantly higher AUC scores compared to state-of-the-art approaches against benchmark anomaly detection datasets, and a multidimensional, multimodal dataset from real climate control devices. Finally, we describe how negative sampling with neural network classifiers have been successfully deployed at large scale to predict failures in real time in over 15,000 climate-control and power meter devices in 145 office buildings within the California Bay Area.
\end{abstract}

\section{Introduction}
\label{introduction}

In this paper we propose a scalable, unsupervised approach to detecting anomalies in data streams of Internet of Things (IoT) connected devices. Complex devices are connected daily and eagerly generate vast streams of multidimensional measurements characterizing their momentary state.  Occasionally, some devices fail, resulting in a system outage, and postmortem analysis reveals that these devices generated unusual states before the outage. Had technical staff been promptly alerted, they could have preemptively fixed the device. It is often impractical or impossible to predict failures on fixed rules or supervised machine learning methods, because failures are too complex, devices are too new to adequately characterize both normal and failure modes in a specific environment, or the environment changes and puts the device into an unpredictable condition. Examples of such complex networked devices include power and climate control in commercial buildings, servers and computers in data centers, badge readers and alarms in physical security systems, and electromechanical components in powerplants. In this paper, we explore an approach that automatically observes a complex system, generates a normal, multidimensional baseline, and detects anomalous measurements from the incoming data stream.

\textbf{Anomaly detection} refers to the problem of finding patterns in data that do not conform to expected behavior \cite{Chandola2009}. Multidimensional data points may be anomalous because, in one or more dimensions, the value lies outside of an expected range, or because an expected correlation between two or more dimensions is violated. For example, suppose a thermostat reports periodic updates of an observed temperature and a setpoint temperature. An anomaly may occur when the observed temperature exceeds the normal range (i.e., too hot or too cold), or when the observed temperature and setpoint temperature values no longer correlate, even if both values themselves fall within normal ranges (i.e., the climate control device fails to maintain a desired setpoint temperature). 

Anomaly detection algorithms generally proceed in two steps. The first step is to characterize a normal baseline model. The second step is to score each new data point as normal or anomalous. This score may be a simple binary label or a real-valued outlier score that quantifies how anomalous the data point is, such as a class probability or confidence \cite{Aggarwal2017} .  

There are several factors that should be considered when developing multidimensional anomaly detection solutions:

\textbf{Noise Dimensions.} IoT devices generate logs or telemetry data streams with high dimensionality, where the anomaly is observable in a subset of the dimensions, but masked in noise dimensions. Unfortunately, with increasing dimensionality, many conventional anomaly detections methods fail to work effectively. 

\textbf{Correlations.} Baseline correlations are multidimensional and may be nonlinear with some dimensions exhibiting strong positive or negative correlations, while others may exhibit weak or no correlations. For example, under normal conditions, the desired setpoint temperature and observed temperature are highly correlated, but the relative humidity and carbon dioxide level may only be weakly correlated.

\textbf{Multimodal.} Oftentimes real-world, normal processes operate in multiple modes. For example, energy efficient thermostats may operate in an eco-mode while the zone is vacant, and comfort-mode while the zone is occupied. These modes may be in distinct regions in feature space, producing none or few intermediate data points from transitions between modes. 

\textbf{Interpretable.} To aid diagnosis and defect remediation, it is often desirable, especially with high-dimensional data, to provide additional information to help understanding the anomaly along with the anomaly score. The anomaly detector should therefore attribute a proportional blame to those variables that contributed to the anomaly score.

In section 2, we present relevant work in anomaly detection, Concentration Phenomenon, and model interpretabilty. Section 3 discusses how negative sampling and binary classifiers can be used for anomaly detection, and applies Integrated Gradients for interpreting anomalies. Section 4 compares performance of our methods against state-of-the-art methods, and demonstrates anomaly interpretability. In section 5, we describe our large-scale implementation to monitor climate control devices and recommend future work.

In summary, this paper makes the following contributions:
\begin{itemize}
    \item a scalable approach to detecting multidimensional outliers, robust under multimodal conditions,
    \item an alternative to one-class anomaly detection using negative sampling, and  
    \item a novel application of Integrated Gradients for anomaly interpretation.
\end{itemize}

\section{Background and Related Work}
\label{background}
We consider anomaly detection to be fundamentally a binary classification task. Each new data point is assigned an anomaly score that indicates how likely it is a member of the normal or anomalous class. However, because comprehensive labeled data are not usually available, the anomalous class is underrepresented, and failure modes cannot be fully characterized, using supervised methods is not straightforward.  Previous works have used various methods to overcome these challenges, including:
(1) negative sampling approaches that explicitly characterize the anomalous space,
(2) one-class classifiers that learn a transformation of the normal data to a characteristic manifold, 
(3) autoencoder-based methods that detect anomalies with large reconstruction errors, and
(4) density-based approaches that identify anomalies as points within sparsely populated regions.
We also review how previous works have applied the Concentration Phenomenon for anomaly detection, and describe relevant works in interpretability. 

\textbf{Negative Selection Algorithms} (NSA) were initially proposed as a biologically-inspired method of detecting computer viruses \cite{Forrest1994}. Most NSAs apply search algorithms that attempt to emulate how antibodies distinguish pathogens from body cells. \cite{Gonzalez2002} first proposed a method that applies NSA to generate a negative sample from an unlabeled positive sample containing both normal and anomalous data, and then trains a classifier on the negative and positive samples to learn a decision boundary between normal and anomalous subspaces. Multiple papers have been written on variations on search-based NSA techniques and evaluating NSA runtime complexity \cite{ji2007,yang2017,hosseini2019}. Because NSA-based anomaly detection relies on many detectors to explore the boundaries of the positive (self) sample, it suffers from large time complexity and space complexity \cite{jinyin2011,ayara2002}. \cite{stibor2005} demonstrated that One-Class SVM outperformed a variable-sized NSA detector algorithm on the Iris-Fisher and biomedical data sets, and questioned whether negative selection is appropriate for anomaly detection.  In this paper we apply negative selection, but propose a much simpler approach with uniform sampling and a binary classifier to build decision boundaries around regions with different sample densities.  

\textbf{One-class classifiers} are trained to learn a transformation function $f:X\rightarrow c$ that generates a scalar value as a class score $c\subset C$, when the input data resembles the observed, and mostly normal, data stream \cite{Aggarwal2017}. However, when $c$ deviates significantly from $C$, $X$ is anomalous. Probably the most prominent one-class classifier is the One-Class SVM (OC-SVM) \cite{schoelkopf2001}. The OC-SVM learns a maximum margin hyperplane that best separates the normal class from the origin given by $W\cdot \Phi (X) -b = 0$, where $\Phi (\cdot)$ is an unknown transformation function, $W$  are its coefficients, and $b$ is a bias. The positive set represents the normal space, and the origin is the only labeled negative instance. The transformation function $\Phi (\cdot)$ is learned indirectly by applying the kernel trick within a Lagrangian dual formulation. Many additional variations to one-class classifiers have been proposed. One Class Neural Networks (OC-NN) \cite{Chalpathy2018} is a deep learning extension to OC-SVM that replaces the SVM with a deep neural network.  Deep Support Vector Data Description (Deep SVDD) \cite{ruff2018} is an extension to the OC-SVM that trains a neural network while minimizing the volume of a lower-dimensional hypersphere that encloses the network’s representation of the data. Anomalous data is likely to fall outside the sphere, and normal data is likely to fall inside the sphere. The soft-boundary Deep SVDD objective consists of a hypersphere volume term, a penalty term incurred by normal points outside the sphere, and a regularization term. The penalty term contains a contamination parameter that controls the trade-off between the volume of the sphere and any violations of the boundary, allowing a configurable fraction of points to be truly anomalous. While Deep SVDD was shown to work using image-based data sets (CIFAR-10, MNIST, and GTSRB), the Deep SVDD objective functions can be readily adapted to multidimensional device state vectors. 

\textbf{Autoencoder and generative methods} use deep learning encoder-decoder architectures, where outliers tend to have greater reconstruction errors than normal points. Deep Autoencoders with Density-Based Clustering (DAE-DBC) \cite{Amarbayasgalan2018} train an auto encoder to minimize the reconstruction error, and then apply density-based clustering (DBSCAN) in the latent space to identify outlier points and clusters. In One Class Classification using Intra-Class Splitting (OC-ICS) \cite{Schlachter2019}, the training data is refined by splitting it into typical and atypical normal sets, using a predesignated threshold on reconstruction error from an autoencoder. Generative adversarial methods, such as AnoGAN \cite{schlegl17} and GANomaly \cite{akcay18}, include discriminators that refine the autoencoder's ability to learn subtle, contextual information and increase the representative power and specificity.

\textbf{Density-based approaches} identify points that are in sparsely populated regions. One widely used approach originally proposed by \cite{Liu2008}, called Isolation Forest, is an ensemble of random trees that recursively partition the data space until all points are isolated. Since anomalies are ‘few and different’, they require fewer partitions on average than the normal points, which are located in dense regions. Consequently, anomalies have shorter path lengths -- the number of edges from the point to the root of the tree -- than normal points.  To compute a point’s anomaly score, the path lengths from all trees are averaged and normalized, and conditioned to return a value of 1 for normal points and 0 for anomalous points. With multimodal datasets, \cite{Harari2018} showed that the Isolation Forest can miss anomalies because it forms false normal regions, proposed an enhancement that partitions the data using hyperplanes with random slopes, called Extended Isolation Forest, and demonstrated improved anomaly detection performance. 

The \textbf{Concentration Phenomenon} describes how manifolds distort as dimensionality increases, and is useful for characterizing anomaly detection algorithms in high-dimensional spaces, but has not been applied extensively in anomaly detection, with a few exceptions. The Concentration Phenomenon was applied previously in a supporting lemma to guarantee convergence for estimating a baseline matrix that is used to quantify the surprise of network user-to-object accesses \cite{Gutflaish2017}. Distance concentration was applied to prove that the proposed outlier factor does not degrade with increasing dimensionality, thereby mitigating the “curse of dimensionality”.

\textbf{Model Interpretability} is an active area of research that seeks an understandable explanation for a model’s decision \cite{doshivelez2017}. Numerous papers have been written about attribution, a type of interpretability that explains what features were most influential in a classifier’s decision \cite{baehrens2009, simonyan2013, shrikumar2017, binder2016}. Integrated Gradients \cite{Sundararajan2017} is an attribution method originally proposed for highlighting important features used by deep neural network classifiers. General approaches to enhancing anomaly detection with interpretability have not been adequately addressed in the literature. However, a few domain-specific implementations have been proposed, such as using Markov Chain methods and cross entropy for identifying attributing anomalous network traffic to multiple correlated traffic flows \cite{nevat2018}.  

\section{Detecting and Interpreting Anomalies}
\label{method}
Consider a target process that generates a sequential stream of multidimensional data points $x(1), x(2),...$, where the $i^{th}$ data point\footnote{For clarity, we drop the sequential index $i: x(i)=x$, and reserve the subscript to index dimensionality.} $x(i)$ , is a $D$-dimensional vector $x=\left \{x_{1}, x_{2},...,x_{D}\right \}$ in $\Re^{D}$. The target process may be either a single discrete unit, or a homogeneous cohort\footnote{In practice, it is often beneficial to group many equivalent devices together into a cohort, because (a) the cohort’s baseline is more robust, and (b) anomaly detection can be performed on multiple devices effectively in parallel within the same execution process.} of equivalent units. Our objectives are to (a) estimate $P(x\in Normal)$, and (b) attribute the anomaly score on to each dimension as an interpretation for each anomalous $x$.

\textbf{Definition 1.} \emph{An \textbf{anomaly} is any data point $x$ with a near zero probability that it was generated by the Normal process: $P(x\in Normal)\approx 0$.}

The \textbf{Normal} process occupies one or more discrete manifolds or volumes of unknown shape, in $\Re^{D}$, which are populated in high density by the target process, except for the small subset of anomalous points that fall outside the Normal volume(s). The \textbf{Anomalous} volume is the complement of the Normal volume in $\Re^{D}$. A \textbf{unimodal process} occupies a single discrete volume, and a \textbf{multimodal process} occupies two or more spatially disconnected volumes.
\subsection{Detecting Anomalies with Negative Sampling}
The Concentration Phenomenon states that there is exponentially more room in higher dimensions than in lower dimensions. Intuitively, as the number of dimensions $D$ increases, the cube has $2^{D}$ corners, so most of the volume is concentrated near them, and manifolds tend to occupy less volume relative to the full space\cite{vanhandel2016,vershynin2018}.  Based on the Concentration Phenomenon, we propose a simple method for developing a labeled data set to train a classifier for our anomaly detection task.

Our approach to anomaly detection is to define two class samples, and train a classifier function $F:\Re^{D}\rightarrow [0,1]$ to distinguish between the two classes. The \textbf{positive class sample}\footnote{Because a positive outcome from a statistical test ordinarily implies the anomalous case, a statistical perspective might argue for reversing the terminology to use \emph{positive} to indicate an Anomalous case and \emph{negative} to indicate a Normal case. However, in this work we use the term \emph{positive} to refer to the space that is sampled by observation and is mostly Normal, and \emph{negative} to refer to an unobserved complement space from which a labeled sample has to be generated.} $U=\left \{ u(1), u(2),...,u(M) \right \}$ is the set of $M$ $D$-dimensional data points\footnote{Every $u$ and $v$ are $D$-dimensional points like $x$, and we use letters $u$ and $v$ to distinguish the positive (observed) sample $U$ from the negative sample $V$.} that were observed by the target process, including a small number of unlabeled, actual anomalies, which our anomaly detection algorithm is expected to detect. Because the positive sample is contaminated with a small number of anomalous data points, there is also a false positive labeling error, $P(u\in Anomalous)>0$. However, because anomalies are rare by definition, the probability that any point drawn from the positive sample is normal is nearly one, independent of dimensionality $D$.  The \textbf{negative class sample} $V=\left \{ v(1), v(2),...,v(N) \right \}$  is chosen independently and uniformly from the cube bounded by the extrema of each dimension plus some small $\pm\delta$, where the volume bounded by the negative sample is strictly greater than the volume bounded by the positive sample, $Vol(U)<Vol(V)$. The \textbf{sample ratio}, $r_{s}=\frac{N}{M}$, governs the ratio between the negative sample size and the positive sample size, where $r_{s}=0$ represents the one-class anomaly detection classifier. Since the negative class is intended to represent the space of anomalies, and there is a nonzero probability that a negative sample point will land within the normal region, $P(v\in Normal)>0$, we also have a false negative labeling error. 

\textbf{Assumption 1} (Sufficiency): \emph{The positive sample U is representative of a stationary, ergodic Normal process.} As with all supervised approaches, the training set should be representative of the prediction set, and it is essential to sample enough to reflect all Normal modes of behavior. In monitoring climate control devices, it is important to sample from all hours of the day, and days of the week, even when fixing seasonal conditions. 

Ideally, we would like to train a binary classifier $F$ with data that has as few labeling errors as possible. Intuitively, it makes sense to develop an algorithm that carefully selects the negative sample to avoid the Normal space. For example, \cite{Gonzalez2002} proposes a type of region-growing approach that avoids choosing points close to the positive sample.  However, such a sampling approach is difficult and/or computationally expensive in high dimensions because we are not able to characterize the positive volume. Instead, we observe that volumes tend to contract in high-dimensional spaces, and propose using uniform i.i.d. sampling for generating the negative sample.

\textbf{Proposition 1.} (Uniform Negative Sampling): \emph{For each dimension $d \leq D$, let $lim_{d}=\left [ min\left ( U_{d} \right ) -\delta, max\left ( U_{d} \right ) +\delta  \right ]$ be a range bounded by the extrema of the positive sample $U$ extended by a conservative positive length $\delta$ that extends $lim_{d}$ beyond the normal space. We assume that the sample size of $U$ is sufficiently large to bound the Normal region. Choose a negative sample $V$, by selecting $N$ points uniformly i.i.d. bounded by $lim_{d}$ for each $d \leq D$. In high dimension, $ D \rightarrow \infty $, false negative sampling error decays exponentially to zero, regardless of the shape of the Normal region.}

\emph{Proof.} While it can be shown that specific shapes such as the sphere or Gaussian contract to zero volume in high dimensions, we choose the hypercube for the normal volume with lengths bounded by the extrema $\Delta u_{d}=max \left ( U_{d} \right ) - min \left ( U_{d} \right )$ since its relative volume decreases most slowly with increasing dimension. Since we are sampling uniformly, the probability of a false positive is the relative volume $P\left ( v \in Normal \right ) = \frac{Vol(V)}{Vol(U)}$.  The length of dimension $d$ in the negative volume is $\Delta v_{d} = \Delta u_{d} + 2 \delta$ . Since $\Delta u_{d} < \Delta v_{d}$ for all $d \leq D$, as the dimensionality $D$ increases, the false negative error descends exponentially to zero $P \left ( v \in Normal \right ) = \lim_{ D  \rightarrow \infty } \prod_{d}^{D}\frac{\Delta v_{d}}{\Delta u_{d} }=0$ .    

The rate at which $P \left ( v \in Normal \right )$ decreases depends on the geometry and volume of the Normal region. Proposition 1 provides upper-bounded asymptotic guarantees that can be strengthened given knowledge of the geometry of the Normal volume. However, when the characteristics of the Normal region are unknown, without loss of generality, we can use Proposition 1 to bound the false negative probability $P \left ( v \in Normal \right ) \leq \prod_{d}^{D}\frac{\Delta v_{d}}{\Delta u_{d} }$ Next, we apply Proposition 1 to develop a simple procedure for developing a dataset that can be used to train an anomaly detection classifier.

\textbf{Proposition 2.} (Labeled Training Set for Anomaly Detection): \emph{Given a sufficiently sampled, high-dimensional dataset from a target process and uniform negative sampling, we can generate a labeled two-class dataset to train a classifier $F$ for detecting anomalies.} 

A good training set requires a low number of labeling errors for a classifier to learn decision boundaries. In this application, both false positive and false negative errors are small. By our definition of an anomaly, false negative occurrences are rare and $P \left ( u \in Anomalous \right ) \approx 0$, so $U$ can be used in training data to represent the Normal class. By Proposition 1, uniform negative sampling can be used to generate accurate labeled data representing the anomalous regions, when $ D  \rightarrow \infty$.  

The \textbf{sampling ratio}, $r_{s}$, specifies the density of negative data points since the Negative Sample volume is fixed. In lower dimensions, it is possible to oversample in the Negative Sample, such that $P \left ( v \in Normal \right ) \approx P \left ( u \in Normal \right )$ and a classifier is unable to learn decision boundaries. The sample ratio should be chosen within $r_{s,min}:P \left ( u \in Anomalous \right ) \ll P \left ( v \in Anomalous \right )$ and $r_{s,max}:P \left ( v \in Normal \right ) \ll P \left ( u \in Normal \right )$.

Given that the classifiers are universal function approximators, such as deep ReLU networks \cite{Hanin2019}, there is no limitation to the number of distinct modes, shapes, or orientations of continuous, high-dimensional Normal volumes. Negative-sampling classifiers detect “bad interactions”, where the values of all dimensions are within Normal ranges, but in aggregate, the points are in the Anomalous region.  

If the positive sample size, the dimensionality, and the classifier hyperparameters (number of estimators, tree depth, number layers, layer width, etc.) are fixed, and we assume that, in general, classifier training time grows linearly with the size of input, then the run-time complexity incurred during training of negative sampling depends on the sampling ratio only, $O(r_{s})$, and remains constant, $O(1)$, at inference.

\subsection{Interpreting Anomalies with Integrated Gradients}
In practice, knowing which dimensions caused an anomaly score helps the user identify the root cause and choose an appropriate fix. In climate control devices, dimensions like zone temperature, carbon dioxide level, humidity, etc., are codependent. Highlighting a single anomalous dimension usually identifies a broken sensor, whereas identifying multiple dimensions helps pinpoint a mechanical failure caused by a defective valve, stuck damper, etc.  As a first step in anomaly interpretation, we would like to quantify a proportional blame for each dimension. Recent work on deep network model interpretability can be applied to variable attribution.  For example, Integrated Gradients \cite{Sundararajan2017}, have been shown to indicate what pixels contributed most to an image classification, or what words contributed to a text classification. The Integrated Gradients method computes and integrates the gradient for each dimension from a neutral baseline point to the observed point. In the image classification task, a black image is commonly used as the neutral baseline, and in the text classification task, a zero-embedding token vector is suitable. However, in anomaly interpretation the neutral baseline is not immediately obvious. Fundamentally, given an anomalous point $x$, we would like to know how to transform $x$ into a suitable normal point $u^{*}$. For example, if the thermostat is generating anomalous data while in eco-mode, we would like to understand which dimensions must be altered to transform the anomalous point to a normal point in eco-mode, rather than in a more remote comfort mode. 

\emph{“A key step in applying Integrated Gradients is to select a good baseline”}\cite{Sundararajan2017}, so we first discuss how to choose a baseline, and then describe how to use Integrated Gradients for anomaly interpretation.

\textbf{Proposition 3.} (Baseline Set for Anomaly Detection) \emph{Points from the positive sample used to train the anomaly detection classifier with high Normal class confidence scores, $U^{*}  \subset U : \forall_{u \in U^{*}} F \left ( x \right ) \geq 1 - \epsilon$ are a sufficient baseline set.}

\emph{Proof.} In the original formulation of Integrated Gradients, $x$ has a high class-confidence score, which requires a baseline point that yields a near-zero class confidence score. In anomaly interpretation it is reversed; since $x$ is an anomaly, $F(x) \approx 0$, the baseline point must be from from the Normal set, $F(u) \approx 1$. By Assumption 1, the positive sample is sufficient and stable. A trained classifier will assign highest Normal class scores to points from regions with the greatest difference in positive and negative sample densities. Tolerance $\epsilon$ depends on the classifier and data, but should be large enough to accumulate enough points to cover all high-confidence Normal regions. Since a uniform distribution guarantees a constant negative sample density, these highest scoring points in $U^{*}$ are in regions with the maximum density of Normal observations.  

Once we have established the neutral baseline set, we then determine how to apply it to any anomalous point. To simplify interpretability, we will choose the single nearest point from $U^{*}$ to anomalous point $x$ as an approximation for the closest point on the surface of the Normal volume.  Since integrated gradients computes the gradient along the straight-line path between the point and the baseline, we chose the chose the baseline point from $U^{*}$ with the minimum Euclidean distance $dist$, $u^{*}=argmin_{u \in U^{*}} \left \{ dist \left ( x, u \right ) \right \}$.  (Using the Euclidean distance assumes a Euclidean space, which requires first normalizing all dimensions.) Along with the proportional blame provided by Integrated Gradients, we have found that the values of $u^{*}$ provide useful interpretation as expected values of the nearest normal. 

Now that we have a baseline point for each anomaly, we can apply the Integrated Gradients Equation (\ref{eq:blame}) to assign a proportional blame, $B_{d} \left ( x \right )$ along the $d^{th}$ dimension for anomaly $x$ and nearest normal baseline $u^{*}$, where $\frac{\partial F \left ( x \right )}{\partial x_{d}}$ is the gradient of the anomaly classifier function $F$, and $\alpha$ is path variable that ranges from 0 at $x$ to 1 at $u^{*}$.

\begin{equation} \label{eq:blame}
B_{d}\left (x \right ) \equiv  \left (u_{d}^{*} - x_{d} \right ) \times \int_{\alpha=0}^{1} \frac{\partial F \left ( x + \alpha \times \left (u^{*}-x \right ) \right )}{\partial x_{d}}d \alpha
\end{equation}

The Completeness Axiom of Integrated Gradients ensures that each dimension is assigned a proportional blame and the sum of the blame across each dimension is bounded by tolerance $\epsilon$ and very close to $1$: $1-\sum_{d \in D} B_{d} \left ( x \right ) \leq \epsilon$.

If we assume the cost for computing the distance and gradient are both constant, then the run-time complexity for applying Integrated Gradients for variable attribution is linear with (a) the number of baseline points $\left | U^{*} \right |$, and (b) the number of $k$ steps to compute the gradient along the path variable $\alpha$ from anomaly $x$ to the nearest baseline point $u^{*}$, $O \left ( \left  | U^{*} \right | \right ) + O \left ( k \right )$.

Putting it all together, this approach is an anomaly detector with a novel method of interpreting the anomaly.  Propositions 1 and 2 describe a method of creating a labeled data set to train an anomaly detection classifier, and Proposition 3 proposes a way of applying Integrated Gradients and a deep network anomaly classifier to interpret an anomaly score. In the following section we will demonstrate the performance compared to other anomaly detection methods with various data sets. 

\section{Experiments}
\label{experiments}
In his section we demonstrate the performance of two negative sampling classifiers against benchmark data tests, a real-world data set from climate control devices, and multimodel, multidimensional synthetic data sets. All results are presented as ROC (i.e., True Positive vs. False Positive Rates) AUC. We also demonstrate how Integrated Gradients can be used to help interpret anomalies. 

As described in the previous section, after normalizing the positive sample, we apply uniform negative sampling with a constant $\delta = 0.05$ but vary the sample ratio, $r_{s}$. The negative sample is assigned class label $0$, and the entire positive sample is labeled $1$ (i.e., no real anomalies are omitted).  

\subsection{Anomaly Detection}
\label{anomaly-detection-experiements}
We present two negative sampling classifiers: one using random forests and the other based on neural networks. The \textbf{Negative Sampling Random Forest (NS-RF)} is parameterized by the number of estimators, the splitting criterion (gini or entropy), maximum tree depth, minimum samples per split and leaf, and maximum number of features per estimator. The \textbf{Negative Sampling Neural Network (NS-NN)} varies the number of hidden layers, where each hidden layer has a dense ReLU sublayer, and a dropout sub-layer. Each layer is configured with the same input width and dropout probability. The output is a sigmoid layer. Training is performed via standard backpropagation using binary cross-entropy loss function. In addition to the number of layers, layer width, and dropout probability, we varied the batch size, the number of epochs, and steps per epoch.

We compared the negative sampling anomaly detectors with two prominent anomaly detectors, and two recent extensions of them. \textbf{One-Class SVM (OC-SVM)}\footnote{We used scikit-learn version 0.22 of OC-SVM} is parameterized by the kernel function (linear, polynomial, RBF, or sigmoid), enabling or disabling the shrinking heuristic, and the contamination factor. Like NS-RF, the \textbf{Isolation Forest (ISO)}\footnote{We used scikit-learn version 0.22 IsolationForest implementation.} is an ensemble-based anomaly detection and is parameterized by the number of estimators, maximum number of samples per estimator, maximum number of features per estimator, and contamination. \textbf{Deep SVDD} is a new deep learning adaptation based on OC-SVM.  In its original version, Deep-SVDD performance was demonstrated exclusively on image-based datasets with a LeNet-type CNN to process the 2D spatial features \cite{ruff2018}. In this study, we replaced the CNN layers with a variable number of dense and dropout layers, but applied the soft boundary and one-class Deep SVDD objective functions. We varied the hidden layer width and dropout probability, the number of epochs, the steps per epoch, contamination, the period to recompute the hypersphere radius, learning rate, decay, and momentum. \textbf{Extended Isolation Forest (EIF)}\footnote{We used the EIF authors’ open source implementation available at:  https://github.com/sahandha/eif.} reduces false positive regions that may occur with Isolation Forest on multimodal datasets by slicing the data along hyperplanes with random slopes\cite{Harari2018}. We vary the number of estimators, the maximum tree depth per estimator.

We selected several familiar benchmark anomaly detection datasets from the Outlier Detection Dataset \cite{Rayana2016}, summarized in Table~\ref{datasets-table}. Additionally, we introduced real-world data from smart buildings, which uniquely exhibit multimodal behavior. When climate control devices fail, they cannot meet the required comfort conditions, and/or consume more energy than fully functional units. The Smart Buildings anomaly dataset\footnote{This data set is not intended to characterize all possible failure modes from climate control devices; the anomaly labels represent only one type of failure mode.} consists of 60,425 multidimensional, multimodal observations derived from 15 Variable Air Volume (VAV) climate control devices collected over 14 days, 8 - 21 October 2019, from office buildings in the California Bay Area. In 1,921 (3.2\%) anomalous observations, the zone air temperatures fall below the zone air heating setpoint or above the zone air cooling setpoint, and are of interest to facilities technicians. In other words, the zone air temperature is normal when it remains above the zone air heating setpoint, and below the zone air cooling setpoint. During working periods, the devices operate in comfort mode with tight constraints between the heating and cooling setpoints. During non-working periods, the setpoints are wider to reduce energy consumption, and hence, there are comfort and eco operating modes. The seven numeric dimensions are: zone air cooling temperature setpoint, zone air heating temperature setpoint, zone air temperature sensor, supply air flowrate sensor, supply air damper percentage command, supply air flowrate setpoint, integer day of week (0-6), integer hour of day (0-23). 

\begin{table}[t]
\caption{Summary of Anomaly Detection Datasets.}
\label{datasets-table}

\begin{center}
\begin{small}
\begin{sc}
\begin{tabular}{lcccr}
\toprule
Data set & Size & Dim & Anomaly \\
\midrule
Forest cover (fc) & 286,048  & 10 & 2,747 (0.9\%) \\
Shuttle (sh)     &  49,097  &  9 & 3,511 (7\%) \\
Mammography (mm)  &  11,183  &  6 &   260 (2.3\%) \\
Mulcross (mc)    & 262,144  &  4 & 26,214 (10\%) \\
Satellite (sa) & 6,435 & 36 & 2,036 (32\%) \\
Smart Buildings (sb)  & 60,425 & 7 & 1,921 (3.2\%) \\
\bottomrule
\end{tabular}
\end{sc}
\end{small}
\end{center}
\vskip -0.1in
\end{table}

Before conducting record trial runs, we performed hyperparameter optimization on AUC for each algorithm, and selected the highest performing parameters. We conducted four formal trials with five-fold cross validation for each dataset and anomaly detector, which generated a total of twenty AUC results per detector algorithm and dataset against a held-out 20\% validation slice. The mean and standard deviation of each dataset-detector combination are presented in Table~\ref{results-table} as percentages. For each of the six detectors, we performed a pairwise Wilcoxon rank-sum test of significance and highlighted top performing algorithms, using a significance threshold of 5\%.  

\begin{table}[t]
\caption{Mean and Standard Deviations of AUC values as \% for benchmark datasets and the Smart Buildings dataset. Highlighted values are the top-scoring detectors based on a 5\% significance threshold.}
\label{results-table}

\begin{center}
\begin{small}
\begin{sc}
\begin{tabular}{lllllll}
\toprule
& ocsvm & dsvdd & iso  & eif  & nsrf & nsnn \\
\midrule
               
fc &53$\pm$20   &69$\pm$7      & 85$\pm$4 &\textbf{93}$\pm$\textbf{1} & 80$\pm$2  & 86$\pm$4 \\
sh &93$\pm$0 &88$\pm$9 &\textbf{96}$\pm$\textbf{1} &91$\pm$1 & \textbf{93}$\pm$\textbf{7}  & \textbf{96}$\pm$\textbf{5} \\

mm     &71$\pm$7     &78$\pm$6        &77$\pm$2  &\textbf{86}$\pm$\textbf{2}  &\textbf{85}$\pm$\textbf{4}   &84$\pm$2    \\
mc        &90$\pm$0     &54$\pm$17        &88$\pm$0  &66$\pm$4 &94$\pm$1    &\textbf{99}$\pm$\textbf{1}     \\
sa &51$\pm$1  &62$\pm$3 &67$\pm$2 &\textbf{71}$\pm$\textbf{3}  &65$\pm$4 &\textbf{73}$\pm$\textbf{3} \\
sb &76$\pm$1     &60$\pm$7        &71$\pm$7  &80$\pm$4  &\textbf{95}$\pm$\textbf{1}    &93$\pm$1   \\ 
\bottomrule
\end{tabular}
\end{sc}
\end{small}
\end{center}
\vskip -0.1in
\end{table}
The selected sampling ratios $r_{s}$ ranged from 30 (Shuttle) to 60 (Satellite). NS-NN architectures ranged from one hidden layer with a width of 64 (Smart Buildings) to 3 hidden layers with a width of 461 (Forest Cover). Performance was better with high dropout rates, ranging between 0.34 (Mammography) to 0.8 (Shuttle). Overall, wide and shallow performed better than deep and narrow architectures. NS-RF configurations ranged between 50 estimators with a maximum depth of 50 (Shuttle) to 150 estimators with a maximum depth of 50 (Smart Buildings). 

We performed supplemental analysis using synthetic data sets to evaluate how negative sampling anomaly detection algorithms perform with 4, 8, 16, and 32 dimensions and 1, 2, and 3 Gaussian modes.  We found that for any dimension, both NS-NN and NS-RF maintained similar performance with each mode, maintained AUCs above the OC-SVM, Deep-SVDD, ISO, and EIF. Even when replacing 25\% of the dimensions with uniform noise, we observed less than 4\% degradation compared to the same number of modes and dimensions with no noise dimensions.

\subsection{Anomaly Interpretation}
\label{anomaly-interpretation-experiments}
We used a synthetic dataset to demonstrate how Integrated Gradients can be used to rank anomalous dimensions and suggest expected normal values. The positive sample consists of 2,500 points drawn from two 16-dimensional Gaussian modes $\mu_{1,2}=\pm2.4\times I_{16}$ and $\Sigma_{1,2}=0.5 \times I_{16}$, with an additional 125 points (5\%) drawn from the uniform distribution to represent true anomalies. Using five-fold cross validation, we trained a NS-NN anomaly detector with two hidden layers, width 64, dropout of 0.1, for 100 epochs, and a sampling ratio $r_{s}=2.0$. All data were normalized before training. After the model is trained, we select a reference baseline $U^*$ by predicting on the training set, and selecting all training points based on $\epsilon = 0.01$.

\begin{figure}[htb]
  \centering
  \begin{minipage}[c]{0.46\columnwidth}
    \includegraphics[width=\columnwidth]{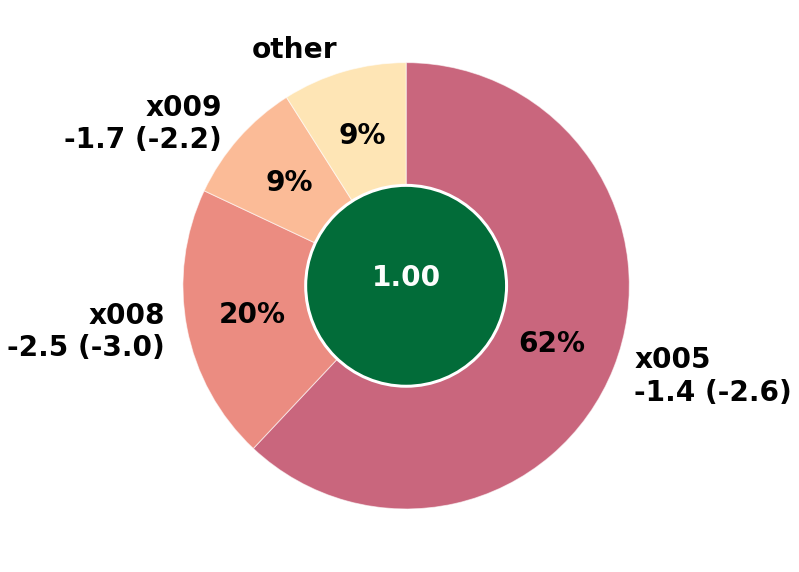}
  \end{minipage}
  \begin{minipage}[c]{0.52\columnwidth}
    \includegraphics[width=\columnwidth]{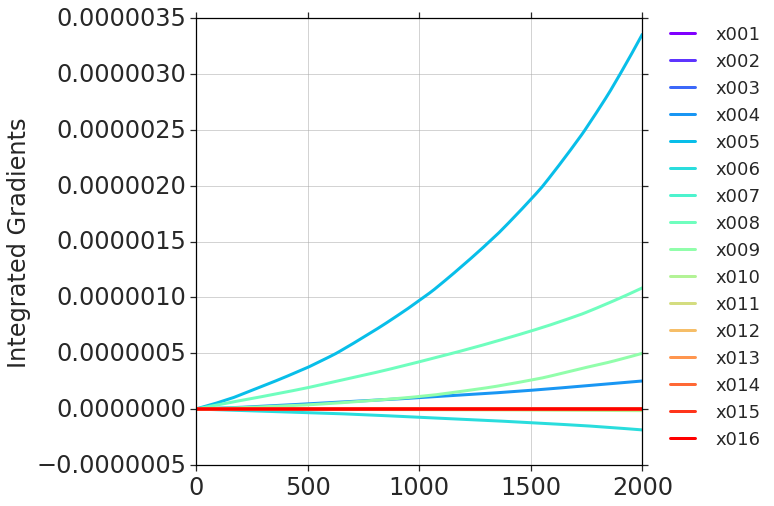}
  \end{minipage}
  \caption{Anomaly Interpretation of a \emph{Normal} point $x$. The left image shows $F(x)=1$ in the center green circle, 
  and the proportional blame $B_{d}$ against dimensions $x005$, $x008$, and $x009$ as exterior wedges. 
  The right chart displays the stepwise integrated gradients from $x$ at $k=0$ to the nearest baseline $u^{*}$ at $k=2,000$. 
  Since the point is normal, the gradients are very small, with $\sum B_{d} \approx 0.$}
  \label{interp-0dim}
\end{figure}

\begin{figure}[htb]
  \centering
  \begin{minipage}[c]{0.46\columnwidth}
    \includegraphics[width=\columnwidth]{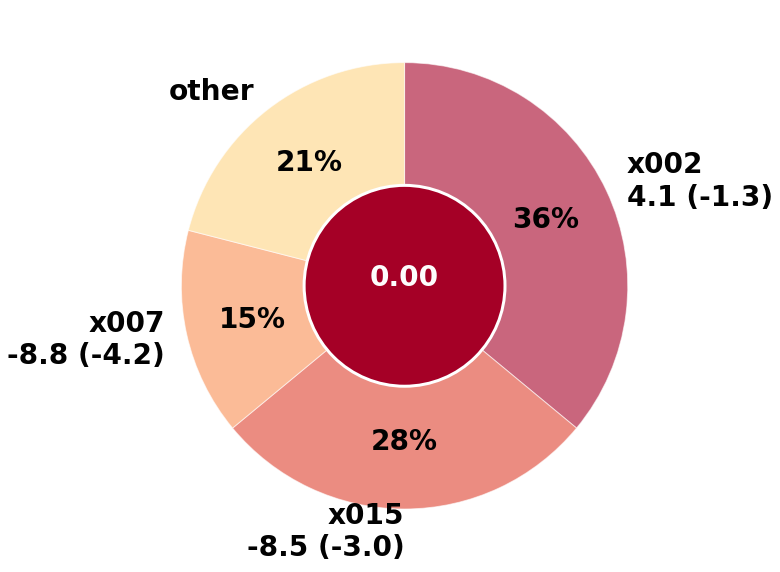}
  \end{minipage}
  \begin{minipage}[c]{0.52\columnwidth}
    \includegraphics[width=\columnwidth]{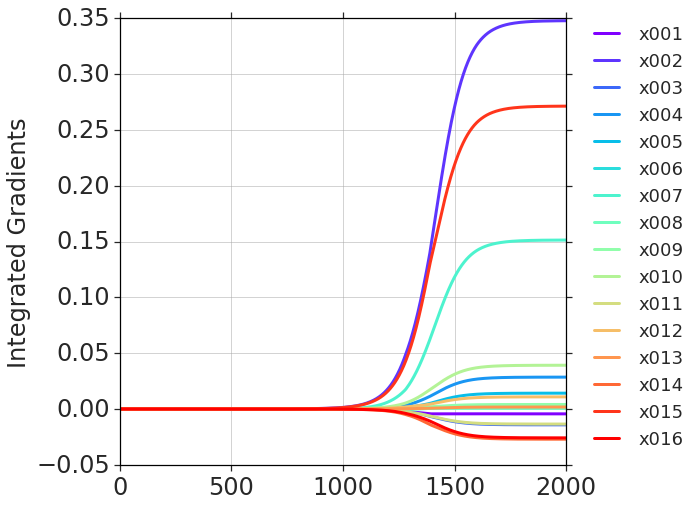}
  \end{minipage}
  \caption{Anomaly Interpretation of an \emph{Anomalous} point $x$ with $F(x) = 0$, 
  Three dimensions ($x002$, $x015$, and $x007$) assigned most of the blame, $\sum B_{d} \approx 1$. 
  The observed and expected normal values, $x_{d}$ ($u_{d}^{*}$), are displayed next to each wedge.  }
  \label{interp-3dim}
\end{figure}

Using the reference baseline set, we use Integrated Gradients to compute the proportional blame on any new point $x$. We selected the one baseline point, $u^{*}$, with the smallest Euclidean distance to $x$. Then, we accumulated the gradients with $k=2,000$ steps along the straight line path from $x$ to $u^{*}$. The Completeness Theorem requires that the gradients will sum to nearly $0$ for Normal points and near $1$ for Anomalous points. In Figure~\ref{interp-0dim}, we first illustrate the response for a \emph{Normal} point, and contrast it with an \emph{Anomalous} point with three anomalous dimensions in Figure~\ref{interp-3dim}. 

\section{Discussion and Conclusion}
\label{conclusion}
It is remarkable that good anomaly detection results are possible with binary classifiers and uniform negative sampling.  It is also interesting that suitable sampling ratios do not appear to grow exponentially with the number of dimensions, making the solution scalable, even to high dimensional spaces. Both Random Forest and Neural Nets generalize well, despite some labeling errors, suggesting that these classifiers are sensitive to the relative sample densities, and generate equivalent results even with different sampling ratios. Because both NS-RF and NS-NN yield fairly similar results, we believe the performance is more associated with the relative sampling densities, than with the type of classifier.

We have applied the Concentration Phenomenon to explain why negative sampling can be combined with classifiers to perform anomaly detection, and show how Integrated Gradients can attribute the anomaly to specific values, even with high-dimensional state vectors. We have demonstrated that negative sampling with random forest or neural network classifiers yield equivalent or higher AUC scores than Isolation Forest, One Class SVM, Deep SVDD, and Extended Isolation Forest against four of five standard benchmark datasets and one multidimensional, multimodal dataset from real climate control devices. 

NS-NN is an integral part of a pilot deployment of our Smart Buildings Fault Detection and Diagnostics (FDD) project. FDD actively monitors over 15,000 power and climate control devices, such as Variable Air Volume (VAV) devices, Fan Coil Units (FCU) and air handlers, boilers, chillers, shade controllers, and electric power meters, etc., installed in 145 office buildings. Because devices in other buildings are periodically added into the platform, it is important that FDD accepts new devices without requiring any manual configuration. Each device reports a multidimensional, numerical state vector in five- to ten-minute intervals with dimensionality ranging from 4 to 20, depending on the device type. All devices are periodically rediscovered and clustered into homogeneous cohorts, and each device cohort is then assigned to its own independent anomaly detection instance. Within a cohort, the dimensionality is fixed; however, the devices in a cohort routinely operate in an occupancy mode during business hours with very strict climate control settings, and an efficiency mode with wider temperature and ventilation tolerances during non-working hours. No comprehensive labeled dataset of failure conditions to train a supervised fault detector is available, and rules-based failure detectors generate an intolerably high false alarm rate. Each NS-NN instance is associated with a single cohort, and periodically retrains a model over sliding historical window to adapt to seasonal changes, and predicts an anomaly score to each new state vector. Persistent anomalous devices are ranked to update a live, enterprise-wide fault detection list. We used Integrated Gradients to help the technicians understand the anomaly by assigning a proportional blame to individual dimensions. The baseline point used by Integrated Gradients for comparison is the nearest normal point observed in the historical training set. The facilities management team reviews daily each of the anomalies and determines which anomalies require a trouble ticket. Over 44\% of all device-level anomalies result in calling technician support. The types of anomalies that generate trouble tickets include stuck airflow dampers and under-ventilated areas, failing sensors, undersized units, unbalanced or uncalibrated units, etc. Most non-actionable anomalies are due to exceptional climate zones, like labs or unoccupied zones. 

There are numerous meaningful directions for future work that could extend negative sampling anomaly detection beyond just numeric features and fixed dimensionality. To increase its applicability and accuracy, negative sampling anomaly detection can be extended to handle categorical dimensions missing values. Many devices generate unstructured or semi-structured textual loglines, and using custom embedding models could be combined with negative sampling anomaly detection. Negative sampling anomaly detection could also be combined with time series analyses to learn and predict sequences of anomaly types, where an anomaly type is localizable to specific region in anomalous space. Some feature dimensions could be readily converted from the time domain into the frequency domain to identify unusual and potentially problematic duty cycles or oscillations. With so many more IoT devices being connected daily in so many different domains, the demand for zero-config, unsupervised anomaly detection will surely continue to provide fertile ground for future research and development.

\section*{Software and Data}
The Python language source code of this work is provided along with the Smart Buildings data set in Multidimensional Anomaly Detection with Interpretability (MADI) at https://github.com/google/madi.

\section*{Acknowledgements}
The author would like to thank the Google Research community and Mukund Sundararajan for instructive and practical advice and for a detailed technical review; Marc Pawliger and the Google Carson Smart Buildings Team for offering the opportunity to apply this approach to its first real-world application; and Rich Dutton and the Google Corp Eng Enterprise AI Team for constructive recommendations and technical discussions. The author would also like to thank the anonymous reviewers for identifying gaps and suggesting improvements.

\bibliography{sipple_icml2020}
\bibliographystyle{icml2020}

\end{document}